# γ-GeSe: a new hexagonal polymorph from group IV–VI monochalcogenides

*Sol Lee,[1,2,‡] Joong-Eon Jung,[1,‡] Han-gyu Kim,[1] Yangjin Lee,[1,2] Je Myoung Park,[3] Jeongsu Jang,[1] Sangho Yoon,[4,5] Arnab Ghosh,[1] Minseol Kim,[1] Joonho Kim,[1] Woongki Na,[3] Jonghwan Kim,[4,5] Hyoung Joon Choi,[1] Hyeonsik Cheong,[3] and Kwanpyo Kim[1,2,*]*

[1]Department of Physics, Yonsei University, Seoul 03722, Korea.

[2]Center for Nanomedicine, Institute for Basic Science (IBS), Seoul 03722, Korea.

[3]Department of Physics, Sogang University, Seoul 04107, Korea.

[4]Department of Materials Science and Engineering, Pohang University of Science and Technology, Pohang 37673, Korea.

[5]Center for Artificial Low Dimensional Electronic Systems, Institute for Basic Science (IBS), Pohang 37673, Korea.

KEYWORDS

Hexagonal GeSe, new polymorph, group IV–VI monochalcogenides, transmission electron microscopy, polarized Raman spectroscopy

ABSTRACT

The family of group IV–VI monochalcogenides has an atomically puckered layered structure, and their atomic bond configuration suggests the possibility for the realization of various polymorphs. Here, we report the synthesis of the first hexagonal polymorph from the family of group IV–VI monochalcogenides, which is conventionally orthorhombic. Recently predicted four-atomic-thick hexagonal GeSe, so-called γ-GeSe, is synthesized and clearly identified by complementary structural characterizations, including elemental analysis, electron diffraction, high-resolution transmission electron microscopy imaging, and polarized Raman spectroscopy. The electrical and optical measurements indicate that synthesized γ-GeSe exhibits high electrical conductivity of $3 \times 10^5$ S/m, which is comparable to those of other two-dimensional layered semimetallic crystals. Moreover, γ-GeSe can be directly grown on h-BN substrates, demonstrating a bottom–up approach for constructing vertical van der Waals heterostructures incorporating γ-GeSe. The newly identified crystal symmetry of γ-GeSe warrants further studies on various physical properties of γ-GeSe.

INTRODUCTION

Two-dimensional (2D) layered crystals have attracted intense research interest owing to their excellent physical, electrical, and chemical properties[1-5] Various properties of 2D crystals strongly depend on the number of layers, interlayer stacking configurations, and types of polymorphs.[1, 6-14] Therefore, exploring the structural configuration space for a given 2D crystal is an important subject in related research fields. In particular, the engineering of 2D polymorphs and their phase transitions has great potential for controlling their electronic properties.[15-18] Moreover, the identification of a new polymorph for a 2D crystal can lead to a better understanding of the material phase boundary between possible structure configurations, opening a new avenue for controlling various properties.

The family of group IV–VI monochalcogenides has recently emerged as a new 2D layered system with interesting properties.[19-23] The group IV–VI monochalcogenides (GeS, GeSe, SnS, SnSe, etc.) usually exhibit a layered puckered structure at room temperatures, similar to black phosphorus (BP). Recent studies have indicated that IV–VI monochalcogenides have great potential in thermoelectric and ferroelectric applications.[24-27] The various synthesis control of IV-VI monochalcogenides was also demonstrated in recent studies, including alloying between different monochalcogenides and the synthesis of in-plane heterostructures.[28-30] The experimentally observed structures of IV–VI chalcogenides include chalcogenide glasses, which are known to have great potential in phase change memory devices.[31, 32] In IV–VI monochalcogenides, individual atoms bond with three nearest neighbors under $sp^3$ hybridization, and their structural configurational space can be quite large. In comparison, various polymorphs have been theoretically predicted for phosphorus cases, and a hexagonal phase of phosphorus (blue phosphorus) has been experimentally reported as a thin-film form grown on a metal substrate.[33-35]

Here, we report the synthesis of the first hexagonal polymorph from IV-VI monochalcogenides, so-called γ-phase GeSe. The structure of γ-GeSe was identified by various complementary characterizations with elemental analysis, electron diffraction, high-resolution transmission electron microscopy (TEM) imaging, and polarized Raman spectroscopy. The identified crystal structure is the first polymorph exhibiting hexagonal crystal symmetry, rather than the conventional orthorhombic phase from IV–VI monochalcogenides. The electrical and optical absorption measurements show that synthesized γ-GeSe exhibits metallic properties, and its conductivity is comparable to that of other 2D layered semimetals. Moreover, γ-GeSe can be directly grown on an h-BN substrate, demonstrating a bottom–up approach for constructing vertical heterostructures incorporating γ-GeSe. The crystal symmetry of the newly identified hexagonal phase of GeSe is distinct from that of conventional α-phase and β-phase GeSe, which warrants further studies on γ-GeSe.

RESULTS AND DISCUSSION

Figure 1 and Table 1 summarize the crystal structures of the experimentally reported GeSe polymorphs. α-GeSe is the most well-studied crystal with a layered puckered structure. This orthorhombic crystal structure is the basic building block for the family of IV–VI monochalcogenides. β-GeSe has recently been synthesized as a bulk crystal form.[36] Although various theoretical studies have indicated the possibility of new types of GeSe polymorphs,[37-41] α-GeSe and β-GeSe have been the only two crystal structures experimentally identified thus far. The four-atom-thick γ-GeSe layer consists of two merged buckled honeycomb lattices (Supporting Information for crystallographic information file) and has recently been predicted by theoretical calculations.[37] In comparison with the phosphorus case, γ-GeSe can be regarded as two merged layers of blue phosphorus.

We synthesized γ-GeSe using a chemical vapor deposition process, as shown in Figure 2a. Polycrystalline α-GeSe powder was used as the source and was heated up to 550 °C (Supporting Figure S1). Vaporized GeSe was recrystallized and collected on a $SiO_2$/Si substrate placed downstream at a temperature of approximately 370 °C. The target substrate was coated with 3-nm-thick Au, which was prepared by thermal evaporation. Au was used as a catalyst for the synthesis of group-IV–VI monochalcogenides.[42-44] We found that the synthesis on the substrate without an Au coating only produced α-GeSe. The crystal often shows a hexagonal shape at the nucleation region, and the corner angle often displays multiples of 60 ° (Supporting Figure S2), which indicates that the synthesized crystal has a hexagonal symmetry. The dagger-shaped flakes were grown from the edges of the hexagons, and their lateral lengths were around 20 μm (Figure 2b, 2c and Supporting Figure S1). SEM images also show the presence of Au nanoparticles at the center of hexagonal-shaped flakes or the tip end of dagger-shaped flakes (Supporting Figure S2), which support that the growth is driven by vapor-liquid-solid (VLS) mechanism via Au catalyst.[45] Atomic force microscopy (AFM) images indicate that the synthesized dagger-shaped GeSe flakes usually exhibit a thickness of approximately 40 nm, as shown in Figure 2d. The crystal surface is quite flat, and the surface roughness measured by AFM is approximately 0.92 nm (Figure 2e).

Raman spectroscopy and TEM analyses of the synthesized product confirmed that the synthesized product is a new GeSe polymorph. Figure 2f compares the Raman spectra acquired from commercially available α-GeSe and the synthesized product, γ-GeSe. The identified Raman peaks (22, 68, 90, 166, 258, and 266 $cm^{-1}$) from γ-GeSe are different from those of α-GeSe. The observed low-frequency Raman shift at 22 $cm^{-1}$ can be assigned to an interlayer shear mode, which suggests that the synthesized product is a layered crystal. The hexagonal phase and composition corresponding to monochalcogenide GeSe was directly confirmed by selected area diffraction (SAED) patterns and energy dispersion X-ray spectroscopy (EDX), respectively. TEM imaging and SAED patterns clearly indicate that synthesized GeSe is a hexagonal crystal with an in-plane lattice parameter of 3.73±0.01 Å, as shown in Figures 2g and 2h. From EDX measurements, we found that the Ge and Se elemental ratio is 1:1 (Figure 2i and Supporting Figure S3). The identified stoichiometric ratio of 1:1 rules out other possible phases of GeSe, such as $GeSe_2$ and $Ge_2Se_3$.[46, 47] We also found that the dagger-shaped flake axis is aligned along the armchair lattice direction of the crystal, as marked with yellow arrows in Figures 2g and 2h.

To identify the crystal structure of the synthesized products, we performed various electron microscopy characterizations. SAED and atomic resolution scanning TEM (STEM) imaging were performed from multiple crystal zone axes, providing complete three-dimensional crystal structural information, as shown in Figure 3 and Supporting Figures S4-S6. Atomic-resolution high-angle annular dark-field (HAADF) STEM images of plan-view γ-GeSe samples show a hexagonal crystal structure with a lattice parameter of 3.73±0.01 Å, as shown in Figure 3b. The SAED patterns acquired by tilting the samples matched well with the simulated SAED patterns from the expected γ-GeSe crystal structure in Figure 1 (Supporting Figure S4). We also prepared cross-sectional TEM samples by a focused-ion-beam (FIB) process (Supporting Figures S5 and S6), which could provide crystal structure information from the flake's side direction.

The layered crystal structure and the stacking sequence were clearly visualized with HAADF-STEM imaging, as shown in Figures 3d-3j. From the [10-10] zone axis, we observed that each individual layer consists of a four-atom-thick unit, Se-Ge-Ge-Se sequence, as demonstrated by atomic resolution STEM imaging and EDX mapping. The atomic sequence can be regarded as A-b-c-A and B-a-c-B under the conventional labeling protocol. Although the atomic numbers of Ge(Z=32) and Se(Z=34) are nearly the same, the HAADF intensity can distinguish between two elements, as shown in Figure 3g. The identified structure is the so-called two merged buckled honeycomb layers,[37] which exhibits $D_{3d}$ symmetry.[48] The identified stacking sequence (A-b-c-A) is unique as other layered crystals with a four-atom-thick layer, including InSe and GaSe, show A-b-b-A[14, 49] sequence with $D_{3h}$ symmetry. We note that the previously reported (SiTe)$_2$ layer in a superlattice shows the same stacking sequence as γ-GeSe.[50] We also identified that γ-GeSe exhibits an AB' stacking sequence, where the second layer is rotated by 180° and stacked on the first layer by one-third of a unit cell shift along the armchair direction of the hexagonal lattice. The identified γ-GeSe has a point group of $C_{6v}$ (6mm) and a space group of $P6_3mc$. By our calculations, we confirmed that AB' stacking is more stable than AB stacking by 26 meV per unit cell.

Polarized Raman spectroscopy was performed to investigate the crystal structure and vibrational properties of γ-GeSe, as shown in Figure 4. The Raman spectrum showed no dependence on the polarization direction when the measurements were performed in the parallel polarization configuration (Supporting Figure S7). This observation is consistent with the $C_{6v}$ (6mm) point group symmetry of γ-GeSe. When we changed the relative angle $\theta$ between the polarization directions of the incident and back-scattered (measured) light, we observed different behaviors, as shown in Figure 4a-d. The Raman peaks at 67 and 164 cm$^{-1}$ did not show any $\theta$-dependence. However, the Raman peaks at 90, 257, and 264 cm$^{-1}$ exhibited the strongest Raman intensity in the parallel configuration ($\theta = 0°$), whereas the Raman signal disappeared in the cross-polarization configuration ($\theta = 90°$). The observed $\theta$-dependences of the Raman intensities are consistent with the group theoretical predictions for the $C_{6v}$ (6mm) point group. We calculated the zone center phonon modes and correlated them with the measured Raman spectra (Supporting Table S1). We found that the peaks at 67 and 164 cm$^{-1}$ correspond to the E modes with horizontal atomic vibrations, and the Raman peaks at 90, 257, and 264 cm$^{-1}$ are assigned to the A modes with vertical vibrational motions (Figures 4e and 4f).

The electrical and optical properties of γ-GeSe, the new polymorph of group-IV–VI monochalcogenides, are of great importance. We fabricated field-effect transistor devices using mechanically transferred γ-GeSe flakes on $SiO_2$/Si wafers. The electrical devices did not show a gate-dependence, and the measured resistances of the devices were quite low (~ 100 Ω). The measured average conductivity of γ-GeSe is approximately $3\times10^5$ S/m, which is comparable to other 2D layered semimetals, such as 1T'-$MoTe_2$[51] and Td-$WTe_2$[52]. The optical absorption of γ-GeSe flakes at near-infrared and mid-infrared regions was also measured using a synchrotron-based measurement setup. Strong absorption below 0.3 eV was observed together with the onset of absorption starting at approximately 0.5 eV (Figure 5c).

First-principles calculations of the electronic band structure of γ-GeSe were performed to understand the observed electrical and optical measurement results. Density functional theory (DFT) in the generalized gradient approximation (GGA) and GW approximation were utilized (See Methods for details). The GGA calculations (Supporting Figure S8) indicate that there is an overlap (0.135 eV) between the valence band and the conduction band, exhibiting a semimetallic behavior. On the other hand, the band structure calculated by the GW approximation showed that γ-GeSe is a semiconductor with an indirect bandgap of 0.33 eV and a direct band gap of 0.54 eV at Γ point (Supporting Figure S8). The calculations based on the GW approximation are regarded as more accurate, whereas the DFT calculations usually underestimate the bandgap size.

From the band structure calculations and experimental measurements, we attributed the observed metallicity of γ-GeSe to the high level of doping in the synthesized samples. We considered two cases of antisite defects with reasonable doping levels (~ $10^{20}$ $cm^{-3}$) estimated from the measured conductivity values. The bandstructure calculations of two anitisite cases ($Ge_{1.01}Se_{0.99}$ and $Ge_{0.99}Se_{1.01}$) under virtual crystal approximation indicate that the high level of either *p*- or *n*-doping can be induced by antisite defects. We found a better agreement with the

*n*-doping case, where the calculated joint density of states (JDOS) captures the important feature in absorption data, the strong absorption below 0.3 eV and the onset above 0.5 eV (Supporting Figure S9). In particular, the absorption feature below 0.3 eV can be attributed to the interband transitions at the conduction bands (Supporting Figure S9). We note that the doping type and detailed doping mechanisms in γ-GeSe are currently under investigation.

2D vertical van der Waals heterostructures have gained wide research interest as novel systems. Using the h-BN flakes as a growth template, we realized vertical γ-GeSe/h-BN heterostructures, as demonstrated in Supporting Figure S10. The multiple dagger-shaped γ-GeSe flakes were nucleated on Au-covered h-BN flakes. The cross-section TEM imaging of γ-GeSe/h-BN samples indicates that Au originally positioned on h-BN flakes was displaced during the growth γ-GeSe, and γ-GeSe/h-BN van der Waals interface without Au insertion could be obtained as shown in Supporting Figure S11. The relative facets between γ-GeSe and underlying h-BN often shows a well-defined geometry, indicating that there is an azimuthal angle correlation between γ-GeSe and h-BN. We found that γ-GeSe and h-BN diffraction patterns are mostly rotationally aligned, as shown in Supporting Figure S10. As-deposited Au on h-BN displays the (111) orientation with a strong azimuthal in-plane correlation with respect to underlying h-BN substrate (Supporting Figure S12), which may play an important role in the observed rotational alignment between γ-GeSe and h-BN.

CONCLUSION

We reported the synthesis and crystal structure of previously predicted γ-phase hexagonal GeSe. The structure of γ-phase GeSe was clearly confirmed by various complementary characterizations, including various TEM characterizations, polarized Raman spectroscopy, and electrical investigations. With a broken inversion symmetry in AB' stacked configuration, γ-GeSe is believed to be ferroelectric. Moreover, from the estimated high doping level (~ $10^{20}$

cm$^{-3}$), γ-GeSe also have high potential in thermoelectric. We expect that our first report on the synthesis of hexagonal crystals in IV–VI monochalcogenides will pave the way for various fundamental studies on the new types of polymorphs and their applications in ferroelectric and thermoelectric applications.

METHODS

**γ-GeSe synthesis:** γ-GeSe was synthesized by a chemical vapor deposition (CVD) process. The growth substrate (SiO$_2$/Si wafers or quartz) was coated with 3-nm-thick Au by thermal evaporation. Polycrystalline GeSe powder (≥ 99.995% purity, Ossila Ltd.) of 5.0 mg was placed in a quartz boat and the growth substrate was placed downstream. After vacuuming a quartz tube with a rotary pump, argon gas (300 sccm) was flowed for 10 min to remove oxygen. While keeping the argon injection, the furnace was heated at 100 °C for 15 min. The rotary pump was turned off, and the pressure of the furnace was maintained at one atmospheric pressure by a leak valve. The furnace was heated to 550 ℃ with Ar gas of 150 sccm (usual heating rate: 15 °C /min) and maintained for 30 min. The target growth substrate was located in the temperature zone of 370 °C during the growth process. After the synthesis, the vacuum pump was turned on, and the furnace was cooled down to room temperature. For γ-GeSe/h-BN heterostructure synthesis, mechanically exfoliated h-BN (2d semiconductors) flakes were prepared on SiO$_2$(300 nm)/Si substrates and coated with 3-nm-thick Au.

**Sample preparations:** TEM samples were prepared on Si$_3$N$_4$ holey TEM grids (Norcada Inc.) by mechanical transfer of γ-GeSe using polydimethylsiloxane (PDMS) support. Cross-section TEM samples were prepared by FIB (crossbeam 540, ZEISS). The nanodevice fabrication was performed using an e-beam lithography system (PIONEER Two, Raith GmbH). Ten-nm-thick Ti and 100-nm-thick Au were deposited by DC sputtering for electrode fabrication. The samples for optical measurement were prepared by mechanical transfer of flakes onto a bare sapphire substrate.

**Characterizations:** TEM imaging, SAED, and EDX were performed using JEOL-2010Plus operated at 200 kV and JEOL double Cs-corrected ARM-200F operated at 200 kV. SEM imaging was performed using JSM IT500HR. The polarized Raman spectra were measured at room temperature by using a home-built micro-Raman system using the 2.81 eV (441.6 nm) line of a He-Cd laser and the 2.41 eV (514.5 nm) line of an Ar ion laser. Raman measurements were performed under vacuum with the laser power below 50 μW. Three volume holographic notch filters (OptiGrate) were used to observe the low-frequency region (<100 cm$^{-1}$). For the polarized Raman measurement, an achromatic half-wave plate was used to rotate the polarization of the linearly polarized incident laser beam to the desired direction in the parallel-polarization configuration. The analyzer angle was set such that photons with a relative polarization angle to the incident polarization passed through. Another achromatic half-wave plate was placed in front of the spectrometer to keep the polarization direction of the signal entering the spectrometer constant with respect to the groove direction of the grating. AFM measurements were performed with Park Systems XE-7 under ambient conditions. Electrical measurements were performed using a parameter analyzer (Keithley 4200A-SCS) under 0.1 ~ 1 mTorr. Electrical conductivity was calculated by $\sigma = (1/R)(L/Wt)$, where $R$, $L$, $W$, and $t$ are the measured resistance, channel length, channel width, and thickness of the γ-GeSe flake, respectively. For absorbance measurements, Fourier transform infrared spectroscopy (FTIR) measurements were performed at the 12D IRS beamline of the Pohang Light Source-II (PLS-II) at room temperature. Absorbance, $A$, was obtained from the reflectance, $R$, and transmittance, $T$, via $A = 1 - (R+T)$. Absorption coefficient, α, was calculated by α = $2.303(A/t)$, where $t$ is the sample thickness.

**First-principles calculations:** We performed density functional theory (DFT) and density functional perturbation theory (DFPT) calculations with the Quantum ESPRESSO[53] code, using the Perdew−Burke−Ernzerhof-type generalized gradient approximation (PBE-GGA)[54] for the exchange-correlation energy functional, norm-conserving pseudopotentials, a kinetic-energy cutoff of 100 Ry for the plane wave, and a 12×12×3 k-point sampling for the integration of the density over the Brillouin zone. The atomic structure of AB'-stacked bulk γ-GeSe was constructed using the experimentally measured lattice constants and was refined by minimizing the total energy of the system with the DFT-D2 scheme[55] with the van der Waals interaction. We calculated the quasiparticle band structure of AB'-stacked bulk γ-GeSe using the GW method implemented in the BerkeleyGW code.[56-58] We used the Godby–Needs generalized plasmon pole model[59, 60] for the frequency-dependence of the inverse dielectric function, a kinetic-energy cutoff of 25 Ry for the dielectric matrix, and a 4×4×1 uniform q-point sampling for the bulk system. We included 2000 bands using the static remainder method.[61] Our parameters for the GW calculation produced converged quasi-particle band energies converged within 0.1 eV. To simulate electron- and hole-doping effects on the band structure of γ-GeSe, we considered antisite defects using the virtual crystal approximation. The frequencies of the Γ-point phonon modes of AB'-stacked bulk γ-GeSe were obtained in the DFPT framework.

**Table 1. Unit-cell information of GeSe polymorphs**

|  | a [Å] | b [Å] | c [Å] | α | β | γ |
|---|---|---|---|---|---|---|
| α-GeSe | 3.83 | 4.39 | 10.84 | 90° | 90° | 90° |
| β-GeSe | 3.83 | 5.81 | 8.09 | 90° | 90° | 90° |
| γ-GeSe | 3.73±0.01 | 3.73±0.01 | 15.4±0.1 | 90° | 90° | 120° |

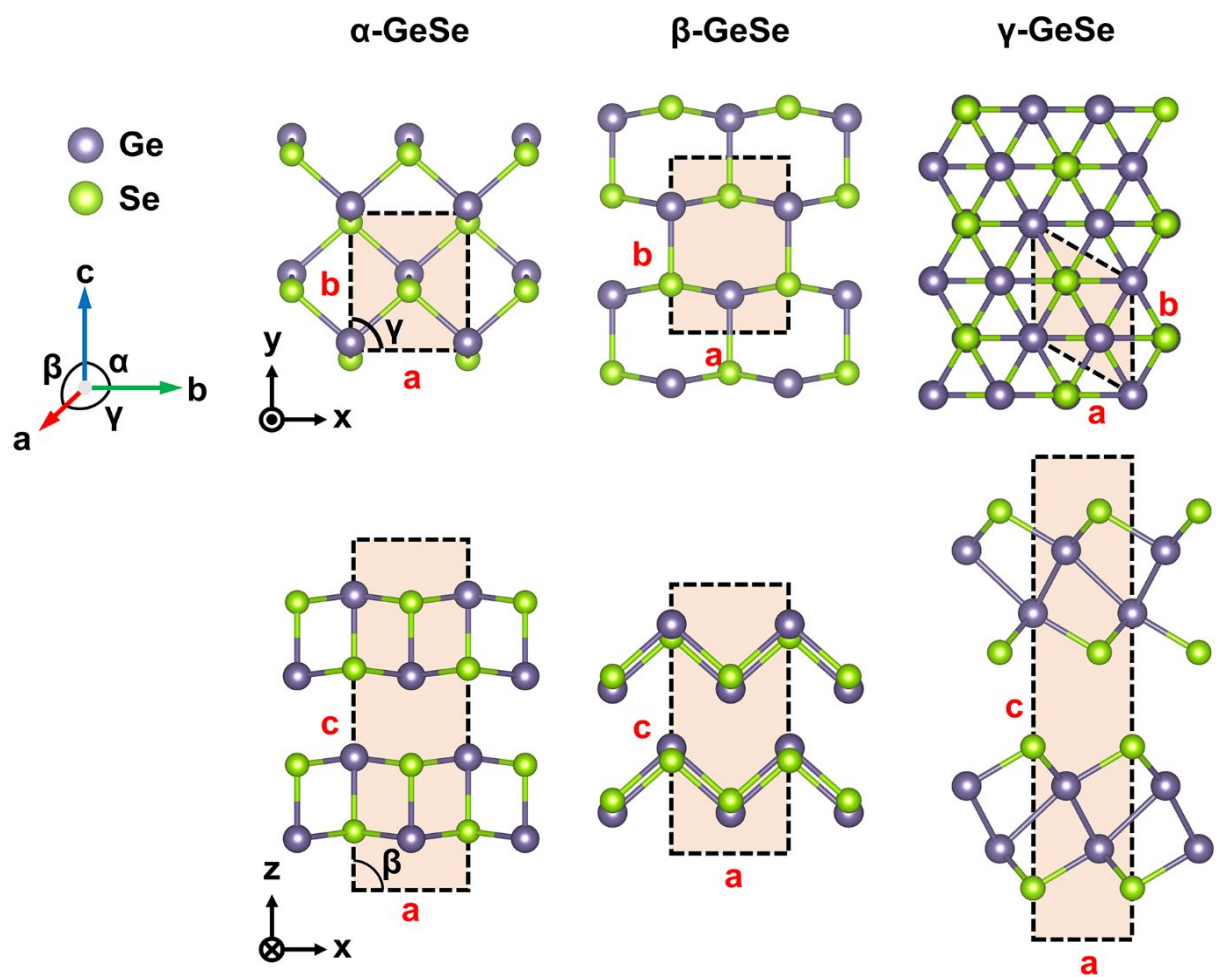

**Figure 1. Crystal structure of GeSe polymorphs.** α-GeSe, β-GeSe, and γ-GeSe are shown from left to right. Germanium atoms, purple; selenium atoms, bright green.

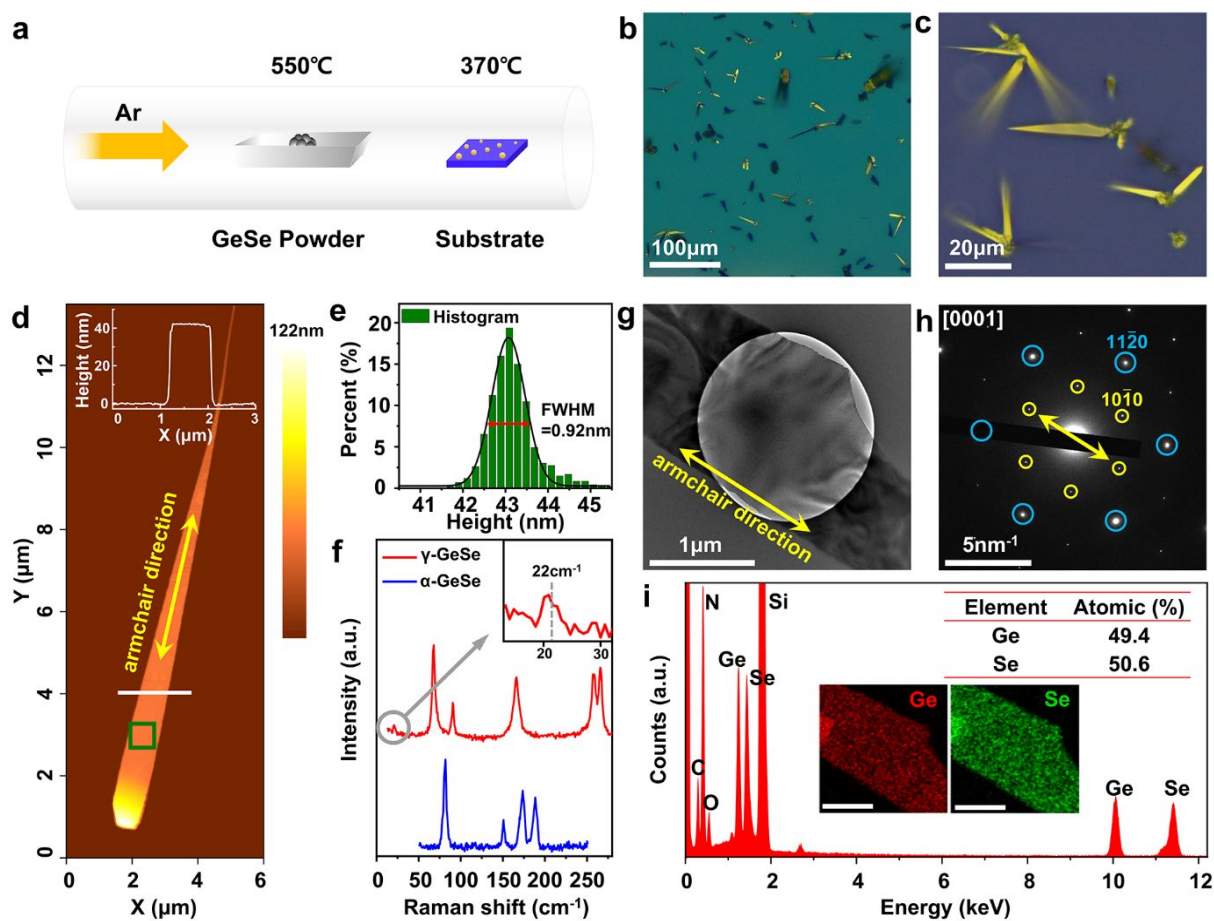

**Figure 2. Synthesis and characterizations of γ-GeSe.** (a) Schematic of PVD process for γ-GeSe synthesis. (b,c) Optical images of dagger-shaped γ-GeSe flakes on SiO$_2$/Si substrate. (d) AFM image of a GeSe flake. The inset is a line profile along the white line in the image. The green square is the field of view for height histogram data for panel e. (e) Histogram of the height distribution measured by AFM. (f) Raman spectra of α-GeSe and synthesized γ-GeSe. (g) TEM image of a GeSe flake. (h) SAED pattern. (i) EDS spectrum obtained from a GeSe flake. The inset images are the EDS mapping images using Ge and Se characteristic peaks.

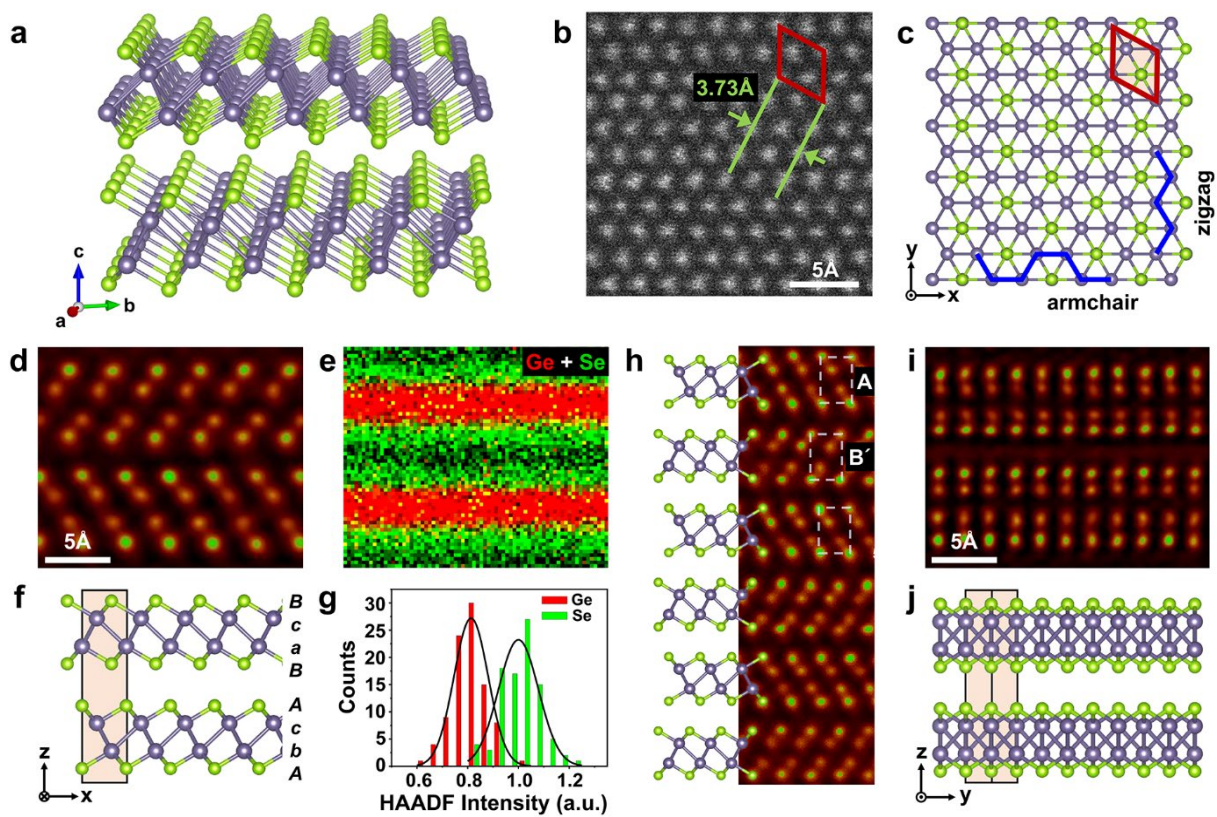

**Figure 3. Identification of γ-GeSe crystal structure**. (a) Schematics of γ-GeSe crystal. (b) Atomic resolution HAADF-STEM image of γ-GeSe along [0001] zone axis and (c) its atomic model. (d) HAADF-STEM image of γ-GeSe along [10-10] zone axis. (e) STEM-EDS mapping data for location shown in panel d. (f) Atomic model of γ-GeSe along [10-10] zone axis. (g) Histogram of measured HAADF intensities of Ge and Se columns. (h) HAADF-STEM image and atomic model with AB' stacking sequence. (i) HAADF-STEM image of γ-GeSe along $[12\bar{3}0]$ zone axis. (j) Corresponding atomic model.

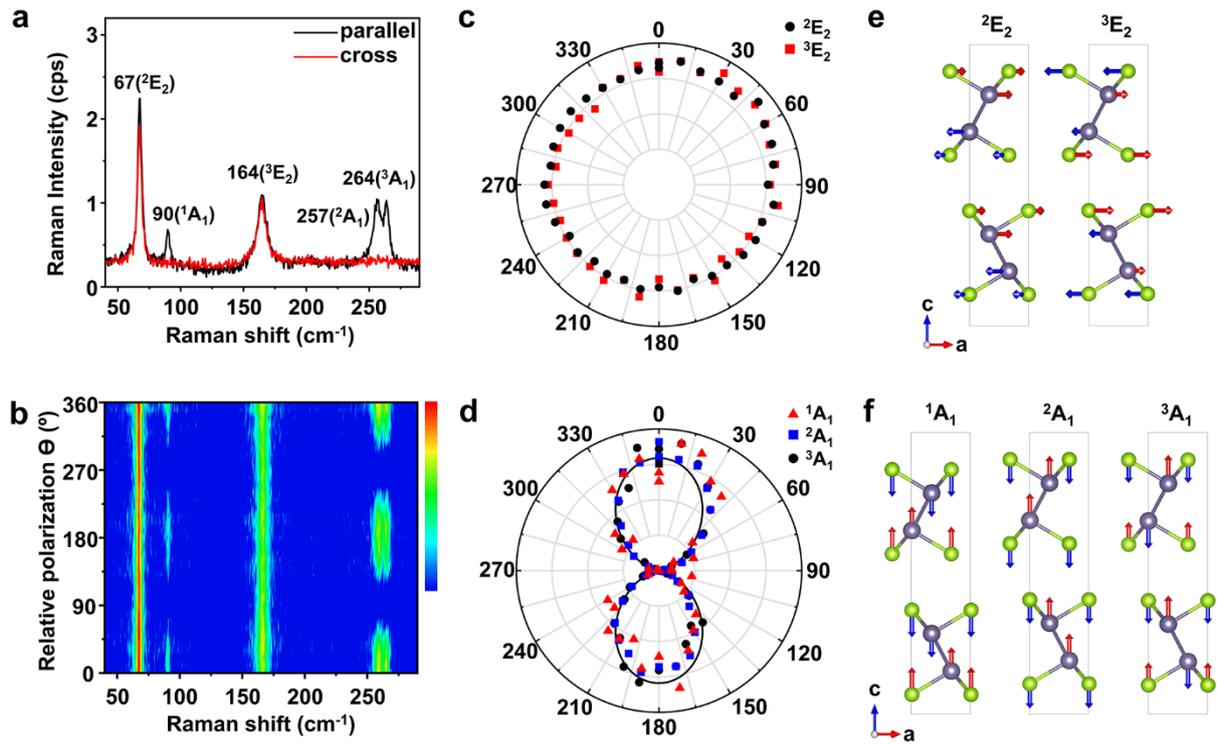

**Figure 4. Polarization-dependent Raman spectroscopy of γ–GeSe.** (a) Raman spectrum of γ-GeSe under parallel and cross polarization measurements. The relative polarization directions between the incident and the scattered light are parallel or perpendicular (cross). (b) Raman intensity plot as a function of the relative angle $\theta$ between the incident and the scattered light. Polar plots of Raman intensities for (c) E modes and (d) A modes as a function of the relative angle θ. The curve is a fit to the $\cos^2\theta$ function. Schematics of (e) E- and (f) A-mode phonon vibrations.

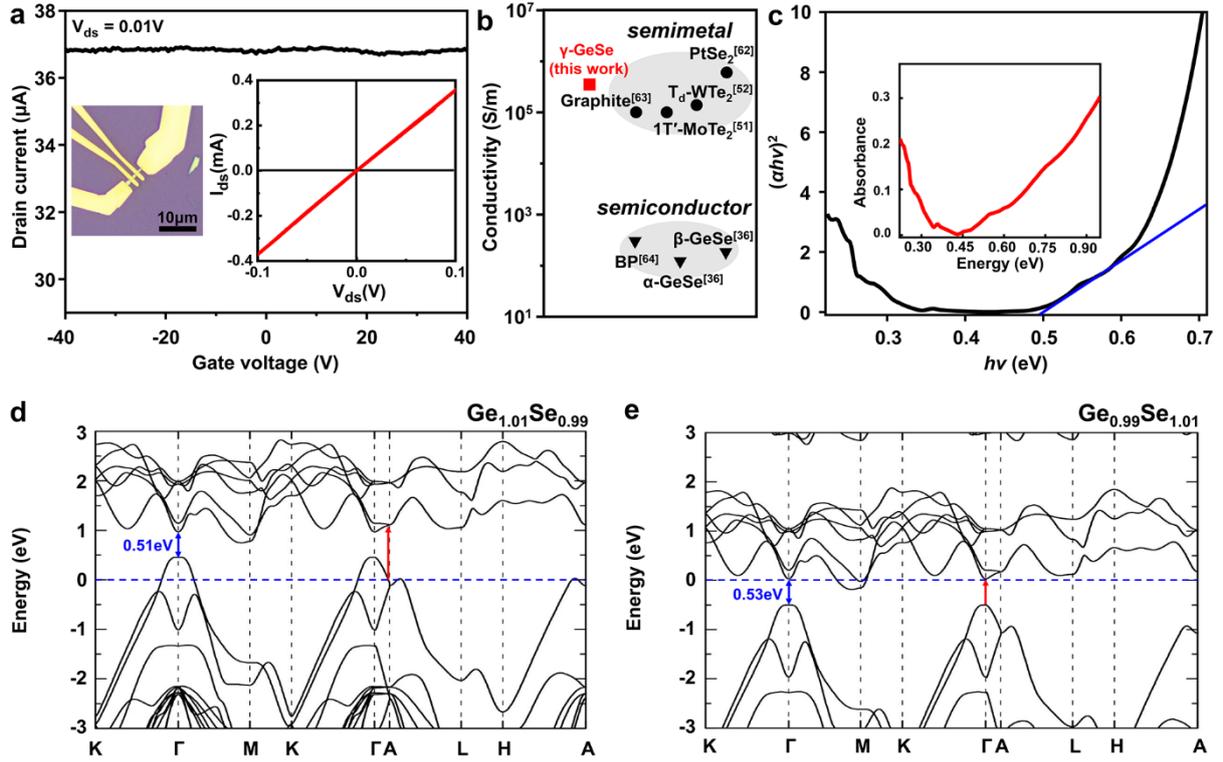

**Figure 5. Electrical and optical properties of γ–GeSe.** (a) Gate-dependence current measurement of γ–GeSe. The inset shows an optical microscope image of a fabricated device and I-V characteristics curve. (b) Electrical conductivity comparison with other 2D layered crystals. [Ref. 36, 51, 52, 62-64] (c) Tauc plot from optical absorbance measurement of γ–GeSe (inset). (d,e) Electrical band structure calculations of γ–GeSe with antisite defects based on virtual crystal approximation.

## ASSOCIATED CONTENT

**Supporting Information**. The Supporting Information is available free of charge on the ACS Publications website.

Extra data on the synthesized products, extra EDX mapping data, extra SAED/STEM analysis data along various zone axes, polarized Raman spectra, and extra electronic bandstructure calculations, synthesis and characterizations of vertical γ–GeSe/hBN heterostructures, Raman shift calculations, and γ–GeSe crystallographic information file.

## AUTHOR INFORMATION

**Corresponding Author**

E-mail: kpkim@yonsei.ac.kr

**Author Contributions**

‡These authors contributed equally to this work.

**Notes**

The authors declare no competing interests.

## ACKNOWLEDGMENT

**Funding Sources**

This work was mainly supported by the Basic Science Research Program of the National Research Foundation of Korea (NRF-2017R1A5A1014862, NRF-2017R1C1B2012729, NRF-

2019R1C1C1003643, 2020R1A2C3013673, and 2019R1A2C3006189), and by the Institute for Basic Science (IBS-R026-D1). Y.L. received support from the Basic Science Research Program at the National Research Foundation of Korea, which was funded by the Ministry of Education (NRF-2020R1A6A3A13060549), and Ministry of Science and ICT (NRF-2021R1C1C2006785). A.G. acknowledges the support from the Yonsei University Research Fund (Post-Doctoral Researcher Supporting Program) of 2019 (2019-12-0033). Computational resources were provided by the KISTI Supercomputing Center (Project No. KSC-2019-CRE-0195). Experiments at PLS-II were supported in part by MSIT and POSTECH.